\documentclass[11pt,twoside]{article}
\usepackage{asp2004}
\usepackage{psfig}
\usepackage{epsf}
\usepackage{graphics}
\usepackage{lscape}

\def\edcomment#1{\iffalse\marginpar{\raggedright\sl#1\/}\else\relax\fi}
\reversemarginpar

\begin{document}
\title{Peculiar Velocities in the Zone of Avoidance: \\
Determining the Distance to the Norma cluster}
\author{Patrick A. {Woudt},$^1$ Anthony P. {Fairall},$^1$ Ren\'ee C. {Kraan-Korteweg},$^2$ John {Lucey},$^3$
Anja C.~ {Schr\"oder},$^4$ David {Burstein},$^5$ and Marshall L. {McCall}$^6$}
\affil{$^1$Dept. of Astronomy, University of Cape Town, Rondebosch 7700, South Africa}
\affil{$^2$Depto. de Astronom\' \i a, Universidad de Guanajuato, Apdo.~Postal 144,
Guanajuato, GTO 36000, M\'exico}
\affil{$^3$Dept. of Physics, University of Durham, South Road, Durham DH1 3LE, United Kingdom}
\affil{$^4$Dept. of Physics \& Astronomy, University of Leicester, University Road, Leicester, LE1 7RH, UK}
\affil{$^5$Dept. of Physics \& Astronomy, Box 1504, Arizona State University, Tempe, AZ, 85287-1504, USA}
\affil{$^6$Dept. of Physics \& Astronomy, York University, 4700 Keele Street, Toronto, ON, M3J 1P3, Canada}

\begin{abstract}
Deep $J$, $H$ and $K_s$ photometry of the low Galactic latitude Norma cluster 
has been obtained with the 1.4-m InfraRed Survey Facility (IRSF) and with SOFI on the NTT. 
Together with spectroscopy taken at the Anglo-Australian Observatory using 2dF, 
the near-infrared Fundamental Plane of the Norma cluster 
will be used to determine the distance to this cluster 
and to assess its location within the Great Attractor overdensity. Potential systematic
uncertainties related to the determination of {peculiar velocities in the Zone of Avoidance}
are discussed in some detail. In particular the effects (uncertainties, systematics and scatter)
of extinction and star-crowding on the derived magnitudes at low latitudes are investigated
in great detail. We present a first look at the emerging $K_s$-band Fundamental Plane 
of the Norma cluster. We expect an uncertainty of $\sim$3\% ($\sim 150$ km s$^{-1}$ at the distance
of the Norma cluster) in our final Fundemental Plane distance estimate based on 76 galaxies.
\end{abstract}

\section{The Norma cluster and the Great Attractor}

The {Norma cluster} ({ACO 3627}, Abell et al.~1989) is a rich and nearby
galaxy cluster at low Galactic latitude ($\ell, b, cz$) =
($325.3^{\circ}, -7.2^{\circ}, 4844$ km s$^{-1}$). Kraan-Korteweg et
al. (1996) and Woudt (1998) showed it to be the nearest massive rich
cluster in the local Universe and comparable in its properties to the
well-studied {Coma cluster}. Were it not for the  obscuration by the
Milky Way (A$_{\rm B}$ $\sim$ 1\fm0) the  prominence of Norma
would have been apparent a century ago and it might have taken the
first place as the best-studied cluster in  the local Universe. Following
the recognition as a nearby rich cluster, however, the Norma cluster has been subjected 
to dedicated multi-wavelength observations:
detailed studies of its X-ray properties have now been made by B\"ohringer et 
al. (1996) and Tamura et al. (1998), the effects of the  intercluster medium 
on the gas contents of spiral galaxies in the  Norma cluster by Vollmer 
et al. (2001), substructure and dynamics by Woudt (1998), and a derivation of its 
optical and near infrared (NIR) luminosity function is in preparation (Woudt et al., in prep.).

A further major point of interest lies in its location at (or close to)  
the bottom of the potential well of the {Great Attractor}. Many years after 
the initial discovery of the GA (Dressler et al.~1987; Lynden-Bell et 
al.~1988) there still remains some ambiguity regarding the extent and 
nature of the GA -- as was evident by various lively discussions at 
this meeting upon the simple question `What is the Great Attractor?'. 

One of our aims in studying the Norma cluster is a highly acccurate distance
determination through a NIR {Fundamental Plane} (FP) analysis with which we
want to pin down the position of the the Norma cluster within the GA 
overdensity, determine its -- and possibly the whole GA's -- movement 
with respect to the Cosmic Microwave Background (CMB), and see whether 
the Shapley Concentration (Scaramella et al.~1989; Allen et al.~1990)
at ($\ell, b, cz$) $\sim$ ($306^{\circ}, +30^{\circ}, 
15\,000$ km s$^{-1}$) exerts a non-negligible attraction on the GA 
(see, e.g., Kocevski et al.~and Lucey, Radburn-Smith \& Hudson, these proceedings).

\section{{Peculiar velocities} in the Zone of Avoidance}

Dedicated galaxy surveys in the ZoA over the last one and a 
half decades across the electromagnetic spectrum -- ranging from 
X-ray (ROSAT, XMM-Newton), optical, near-infrared (2MASS, DENIS), 
far-infrared (IRAS) to radio wavelengths (HIPASS)\footnote{The 
reader is referred to the three dedicated conference volumes on 
galaxy surveys in the Zone of Avoidance for a comprehensive 
overview of the various surveys (ASP Conf. Ser., Vol. 67 (1994), 
Vol. 218 (2000) and this volume).} -- have led to an almost
complete reduction of the ZoA in the distribution of galaxies, even in 
redshift-space (e.g., Henning, Kraan-Korteweg \& Staveley-Smith, and 
Koribalski, these proceedings).

Despite these advances, a broad Zone of Avoidance in peculiar velocity data
has remained at low Galactic latitudes. This is mostly due 
to the difficulty in determining the Galactic foreground extinction 
to the desired level of accuracy. In addition, 
it is not well understood (quantitatively) what the effects are of 
star-crowding on photometry, i.e., how accurately one can subtract the
foreground stars from the underlying galaxy? Both effects (extinction and 
star-crowding) could lead to large systematic uncertainties in peculiar 
velocities derived from, e.g., surface brightness fluctuation, Tully-Fisher, 
and FP analyses. 
An early example of an extremely high peculiar velocity measurement 
for the Norma cluster ($v_{pec} = 1760 \pm 355$ km s$^{-1}$) 
using the $I$-band Tully-Fisher relation had its origin in low-number
statistics as well as large systematic photometric 
errors (Mould et al.~1991; see discussion in Woudt 1998).


Several important structures in the local Universe lie at low Galactic latitude,
e.g. the GA, the {Perseus-Pisces Supercluster} and 
the {IC342/Maffei group} (Buta \& McCall 1999). 
In order to assess their significance in terms of the velocity 
field generated by them, it is necessary to obtain reliable distance 
estimates to the cores of these overdensities. 
The application of the FP requires reliable measurements of photometric parameters, i.e.
$\log (r_e)$ and $\mu_e$. At low Galactic latitude the two key issues
are the sizable and uncertain Galactic extinction and the large level
of foreground stellar contamination. It is therefore 
important to study extinction and star-crowding 
(Buta \& McCall 1999) quantitatively, at least for groups and clusters 
subjected to a modest amount of extinction in high galaxy-density 
regions (e.g., the GA region). Prime examples of such clusters are 
the Norma cluster (Kraan-Korteweg et al.~1996; this paper) and 
{CIZA\,J1324.6-5736} (Ebeling, Mullis \& Tully 2002; Mullis et al., these proceedings).

\section{The near-infrared Fundamental Plane of the Norma cluster} 

We have selected a sample of 76 early-type galaxies in the Norma cluster 
from which the $R_C$, $J$, $H$ and $K_s$ FP will be constructed. 
Spectroscopy for these galaxies has been obtained using the {2dF spectrograph}
at the Anglo Australian Observatory. 
One of the main advantages of using the near-infrared FP for low-latitude 
clusters is that the colour information of the observed elliptical 
galaxies allows a direct measure of the foreground extinction 
(see Sect. 3.3) which, within errors, is not sensitive to 
the adopted Galactic extinction 
law (e.g., Cardelli, Clayton \& Mathis 1989). 

With our large sample of 76 galaxies we expect to be able to determine 
the distance of the Norma cluster to a few per cent -- in the absence 
of systematic uncertainties -- given that the distance to an individual
galaxy using the near-infrared FP is accurate to $\sim$18\% (Mobasher 
et al.~1999).

\subsection{Near-infrared photometry with SIRIUS and SOFI}

We have obtained near-infrared photometric observations of the Norma 
cluster at two different observing sites using two complementary approaches, 
namely pointed observations (in the $K_s$-band) of selected early-type 
galaxies using {SOFI on the NTT} within the Abell radius (1.75$^{\circ}$) 
of the Norma cluster (4 nights in June 2000), and a simultaneous $J$, $H$ 
and $K_s$-band survey of the central 1/3 Abell radius ($\sim 0.6^{\circ}$) 
of the Norma cluster using {SIRIUS on the InfraRed Survey Facility} (IRSF; 
Nagayama et al.~2003) at the Sutherland site of the SAAO (three weeks in 
June/July 2001 and July 2002).  A detailed description of the SOFI and 
IRSF observations will be presented elsewhere. Here we present a 
preliminary analysis of these data, demonstrating the feasibility of 
obtaining accurate photometry at low Galactic latitude.

Despite the availability of 2MASS photometry (Jarrett et al.~2000)
for most of the galaxies in our sample, the 
level of accuracy needed in determining peculiar velocities
-- particularly in the presence of high stellar 
densities (see Sect. 3.2) -- requires deeper observations at high 
resolution (i.e., small pixel scales). 
Both SOFI and SIRIUS meet those requirements, with pixel scales of 0.29 
and 0.45 arcsec/pixel, respectively, compared to the 2.0 arcsec/pixel 
of 2MASS. Table~\ref{woudttab1} gives an overview of the main 
characteristics of our SIRIUS and SOFI observations, in comparison 
to 2MASS.

\begin{table}[!ht]
\caption{An overview of the photometric observations}
\smallskip
\begin{center}
{\small
\begin{tabular}{lccc}
\tableline
\noalign{\smallskip}
   & SOFI  & IRSF & 2MASS \\
\noalign{\smallskip}
\tableline
\noalign{\smallskip}
Telescope size& 3.6 m & 1.4 m & 1.3 m \\
Filters   & $K_s$ & $J$, $H$, $K_s$ & $J$, $H$, $K_s$\\
Field of view & $4.9' \times 4.9'$ & $7.8' \times 7.8'$ & $8.5' \times 8.5'$\\
Pixel scale  & $0.29''$/pix & $0.45''$/pix & $2.0''$/pix \\
Exp.time (total) & 300 s  & 600 s  & 7.8 s\\
Mean FWHM &  $1.08''$ & $1.35''$  &  $\sim 2.9''$   \\
\noalign{\smallskip}
\tableline
\end{tabular}\newline
\footnotesize{Note: Both the IRSF and 2MASS have simultaneous $J$, $H$ and $K_s$ imaging capabilities.}}
\end{center}
\label{woudttab1}
\end{table}

Standard calibration was performed using the near-infrared standard 
stars of Persson et al.~(1998). In addition, a large number of 
`calibration galaxies' were observed (with $K_s^{20}$ magnitudes 
in the range of $\sim 7 - 13$ mag), primarily for the purpose of 
obtaining uniformity with other peculiar velocity surveys (e.g., 
Pahre 1999; Jones et al., these proceedings; Hudson, these proceedings). Apart from the obvious calibration issues, 
these galaxies were also used to test the effect of star-crowding 
on surface photometry by artificially adding star fields (observed 
around the Norma cluster) to high Galactic latitude galaxies 
(see Sect.~3.2).

Standard data reduction and analysis were all performed in 
{IRAF}\footnote{IRAF is distributed by the National Optical
Astronomy Observatories, which are operated by the Association of 
Universities for Research in Astronomy, Inc., under cooperative 
agreement with the National Science Foundation.}. In particular, 
we used the ELLIPSE task in IRAF to determine the surface 
brightness profiles for each galaxy. Both our IRSF and SOFI 
photometry of the standard galaxies compare very well with 
2MASS (see the top half of Table~\ref{woudttab2}); the IRSF, 
SOFI and 2MASS data 
are all on the same photometric system. For this initial comparison, we have 
used isophotal magnitudes at the radius of the $K_s = 20.0$ 
mag/arcsec$^2$ isophote as given by 2MASS.

\begin{table}[!ht]
\caption{Photometric comparison between SOFI/IRSF and 2MASS}
\smallskip
\begin{center}
{\small
\baselineskip 15pt
\begin{tabular}{lccl}
\tableline
\noalign{\smallskip}
\multicolumn{4}{c}{\bf Calibration galaxies}\\
\noalign{\smallskip}
\tableline
\noalign{\smallskip}
   & $\Delta m$ (mag) & No. galaxies & \\
\noalign{\smallskip}
\tableline
\noalign{\smallskip}
$\Delta J^{20}$         & $+0.006 \pm 0.011$ & 10 & (2MASS -- IRSF)\\
$\Delta K^{20}_{s}$     & $-0.004 \pm 0.018$ & 10 & (2MASS -- IRSF)\\
$\Delta K^{20}_{s}$     & $-0.004 \pm 0.009$ & 21 & (2MASS -- SOFI) \\
\noalign{\smallskip}
\tableline
\noalign{\smallskip}
\multicolumn{4}{c}{\bf Galaxies in the Norma cluster}\\
\noalign{\smallskip}
\tableline
\noalign{\smallskip}
   & $\Delta m$ (mag) & No. galaxies & \\
\noalign{\smallskip}
\tableline
\noalign{\smallskip}
$\Delta J^{20}$         & $+0.027 \pm 0.041$ & 17 & (2MASS -- IRSF)\\
$\Delta K^{20}_{s}$     & $+0.046 \pm 0.028$ & 18 & (2MASS -- IRSF)\\
$\Delta K^{20}_{s}$     & $+0.060 \pm 0.026$ & 12 & (2MASS -- SOFI) \\
\noalign{\smallskip}
\tableline
\noalign{\smallskip}
$\Delta K^{20}_{s}$     & $+0.017 \pm 0.014$ & 8 & (IRSF -- SOFI) \\
\noalign{\smallskip}
\tableline
\end{tabular}
}
\end{center}
\label{woudttab2}
\end{table}

\subsection{{Star-crowding}}

In analysing the low Galactic latitude IC\,342/Maffei group 
of galaxies, Buta \& McCall (1999) have developed a routine 
(dubbed KILLALL) within the IRAF image reduction package for 
efficient star removal.

We have applied {{KILLALL}} to all our galaxies, albeit slightly 
modified; instead of letting KILLALL take out the galaxies by 
subtracting a median-filtered image, in our first iteration we 
model and subtract the galaxy (with ELLIPSE and BMODEL, using 
sigma-clipping to mask out most of the star light). During the 
second iteration -- with the stars now well-determined and 
subtracted -- an improved model of the galaxy is made (again 
with ELLIPSE and BMODEL, but this time no sigma-clipping is 
applied) and subtracted before running KILLALL a second time. 
A further iteration is possible if deemed necessary, but often 
two iterations are enough.

\begin{figure}[t]
\plotfiddle{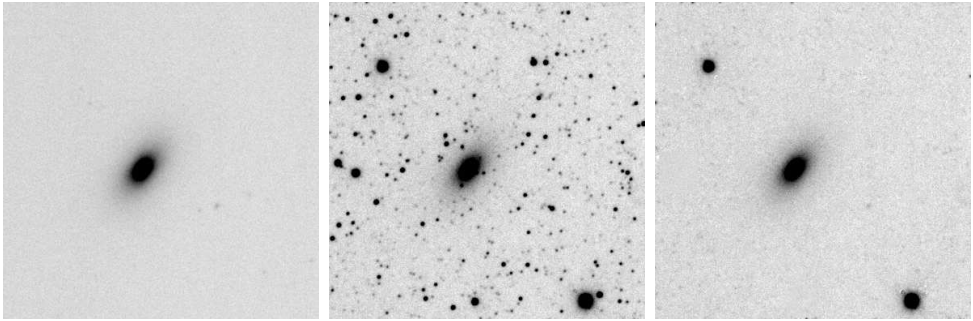}{4.1cm}{0}{70}{70}{-205}{-240}
\caption{Simulations of the effects of star-crowding on surface 
photometry. The left panel shows the original ($K_s$) image of 
NGC 4387 taken with SIRIUS at the IRSF. The middle panel shows 
the same image, but now with a typical star field in the Norma 
cluster region (star field \#49, see Table~\ref{woudttab3}) 
superimposed. The right panel shows the same field, after 
stars were subtracted using KILLALL. The two remaining stars in
this panel are saturated, but are far enough away from the galaxy.}
\label{woudtf1}
\end{figure}

In order to test if KILLALL has removed all the superimposed foreground star 
light, we have added a stellar density similar to that found in the Norma cluster 
region to our calibration galaxies with well-determined magnitudes in order to 
test our procedures. 
At this point, the `calibration galaxies' appear indistinguishable 
from the real Norma cluster galaxies of our selected sample of 
76 early-type galaxies. The modified KILLALL routine is then 
applied to the star-added images and the resulting $K_s^{20}$ 
magnitude compared to the original value. 
An example (for {NGC 4387}) is shown in Figure~\ref{woudtf1}.

\begin{table}[!ht]
\caption{Results of the star subtraction simulation}
\smallskip
\begin{center}
{\small
\baselineskip 15pt
\begin{tabular}{lccccc}
\tableline
\noalign{\smallskip}
   & $\ell$ & $b$ & $\Delta K^{20}_s$ (mag) & FWHM & No. gal. \\
\noalign{\smallskip}
\tableline
\noalign{\smallskip}
Star field \#03    & $325.3^{\circ}$ & $-7.1^{\circ}$  & $+0.008 \pm 0.010$ & $1.24''$ & 9 \\
Star field \#49    & $325.5^{\circ}$ & $-6.8^{\circ}$  & $+0.004 \pm 0.007$ & $1.35''$ & 9 \\
Star field \#54    & $325.0^{\circ}$ & $-6.9^{\circ}$  & $-0.009 \pm 0.011$ & $1.13''$ & 9 \\
Star field \#68    & $325.6^{\circ}$ & $-7.5^{\circ}$  & $-0.000 \pm 0.006$ & $1.35''$ & 9 \\
Star field \#89    & $324.9^{\circ}$ & $-7.6^{\circ}$  & $-0.015 \pm 0.007$ & $1.24''$ & 9 \\
\noalign{\smallskip}
\tableline
\end{tabular}
}
\end{center}
\label{woudttab3}
\end{table}

\begin{figure}[!ht]
\plotfiddle{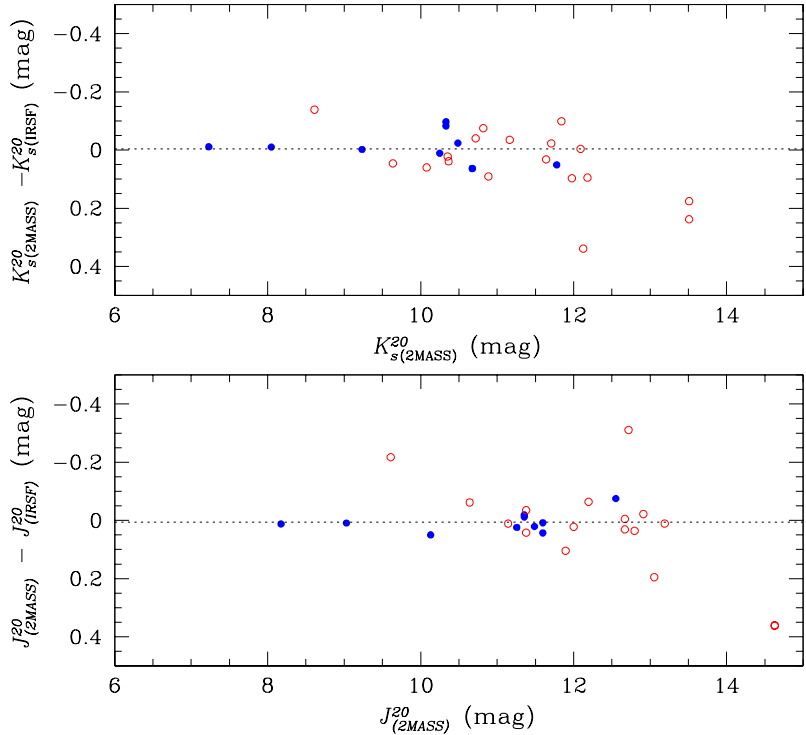}{9.5cm}{0}{57}{57}{-185}{-110}
\caption{A comparison between IRSF and 2MASS phometry in the 
$K_s$-band (upper panel) and $J$-band (lower panel). Our calibration
galaxies are indicated by the filled circles and show good agreement. 
The galaxies in the Norma cluster (open circles) 
at $b \sim -7^{\circ}$ show a much larger scatter.}
\label{woudtf2}
\end{figure}

For each of the nine calibration galaxies, five different star 
fields across the Norma cluster were used and the mean offsets 
(original $K_s^{20}$ $-$ resulting $K_s^{20}$ magnitude following 
the KILLALL procedure) were calculated.  The results of this 
procedure are given in Table~\ref{woudttab3}. It is clear that 
for our high quality near-infrared data of the Norma cluster, 
contamination by foreground stars has a negligible effect 
on the surface photometry ($\la 0.01$ mag on average).

As mentioned in Section 2, 2MASS photometry is available for a 
large number of galaxies in the Norma cluster, although at a
Galactic latitude of $b \sim -7^{\circ}$, the 2MASS extended 
source catalogue (2MASS XSC) is not complete. The prime 
reason for this is stellar confusion (see also Kraan-Korteweg \& Jarrett,
these proceedings). The stellar crowding also results in a much larger scatter 
in the 2MASS photometry at low Galactic latitude, primarily due 
to the limited angular resolution of 2MASS. In Figure~\ref{woudtf2}, 
we show the comparison between 2MASS and IRSF photometry in 
the $K_s$-band (upper panel) and $J$-band (lower panel) for 
both the calibration galaxies (filled circles) and the Norma 
cluster galaxies (open circles). Even fairly bright galaxies 
in the Norma cluster can differ by $\sim 0.15-0.2$ mag, and 
for galaxies fainter than $K^{20}_s \approx 12$ mag and 
$J^{20} \approx 12.5$ mag the scatter increases rapidly. 

The lower half of Table~\ref{woudttab2} shows the mean offsets 
between 2MASS and the IRSF/SOFI photometry for the galaxies 
in the Norma cluster. At first sight, 2MASS appears to be too 
faint by $K^{20}_s \sim 0.05$ mag on average for the low-latitude 
galaxies. If this trend is confirmed by the full sample of 
76 galaxies, it would imply a potential source of systematic 
uncertainty in the FP distance if low-latitude 2MASS photometry 
is to be compared with 2MASS high-latitude photometry. 
The internal comparison (IRSF-SOFI) shows a much smaller offset 
(based on only 8 galaxies!); the smaller offset is consistent 
with the results of the star-adding simulation.

\subsection{Determining the {Galactic foreground extinction}}

Burstein (2003) has given an extensive overview of the advantages
and disadvantages of the use of the Burstein-Heiles (BH; 
Burstein \& Heiles 1978) method and the DIRBE/IRAS (IR) reddening 
maps (Schlegel, Finkbeiner \& Davis 1998) for predicting Galactic 
foreground reddening along the line-of-sight. 

The Norma cluster lies uncomfortably close to the region 
$230^{\circ} < \ell < 310^{\circ}$ and $|b| < 20^{\circ}$ where 
the BH method predicts reddenings a factor 2 too high due to local
variations in the gas-to-dust ratio (Lynden-Bell et al.~1988). 
In this paper we therefore only use the IR reddening predictions.
As pointed out by Burstein (these proceedings), the limiting factor with any of
these methods (BH and IR) is the angular resolution of the survey
in question, given the patchiness of the Galactic dust distribution. 
The resolution of the Schlegel et al.~IR reddening maps is set by the 
$6'$-pixels of the IRAS survey. This will lead to an observable bias
if one is using individual elliptical galaxies (typically with diameters
of $\la 1'$, ignoring variations of the foreground reddening across the
galaxy) to determine the line-of-sight reddening; both BH
and IR methods tend to overpredict the reddening in high extinction
areas, as the galaxies found in galaxy surveys are preferentially found
in regions of smaller extinction. This effect is clearly illustrated by
Nagayama et al.~(2004) for line-of-sight reddenings towards 
elliptical galaxies in the PKS\,1343$-$601 cluster at $(\ell, b, cz) 
\sim (309.7^{\circ}, +1.7^{\circ}, 3872$ km s$^{-1}$).

Despite the relatively low (averaged) value of $E(B-V) \approx 0.2$ mag for 
the Norma cluster, we cannot rely solely on the IR reddening maps; 
line-of-sight reddenings will need to be determined based on the 
observed (i.e., reddened) near-infrared colours for each of the 76 
early-type galaxies in our sample in order to locally calibrate the IR 
reddening maps.

\begin{figure}[t]
\plotfiddle{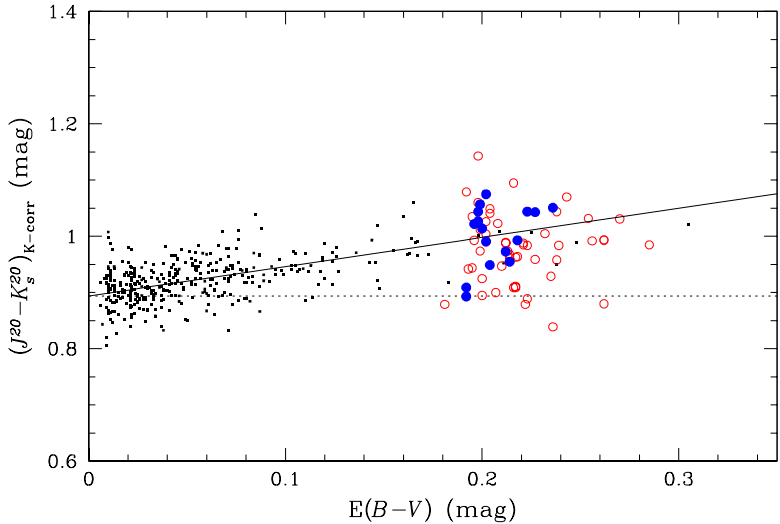}{7.0cm}{0}{57}{57}{-185}{-110}
\caption{The observed k-corrected ($J^{20}-K^{20}_s$) colour of elliptical galaxies. 
The small dots correspond to 380 of the 402 
giant elliptical galaxies from Burstein (2003) for which 2MASS 
photometry is available. The big filled circles show galaxies
in the Norma cluster with IRSF photometry, and the open circles 
are galaxies in our FP sample with 2MASS photometry.
The outlying galaxy with IRSF photometry corresponds to WKK\,6299, 
which is located in the halo of a bright star and will
be excluded from our final FP sample. The dotted horizontal line
marks $(J^{20}-K^{20}_s)^0 = 0.894$ mag (see text) and the solid line
equals $E(B-V) = 0.52 E(J-K_s)$ (and is {\em not} a fit to the data).}
\label{woudtf3}
\end{figure}

In Figure~\ref{woudtf3} we show the observed (k-corrected) $J^{20}-K^{20}_s$ colour 
of Burstein's (2003) sample of giant ellipticals (gE) as a function of $E(B-V)$ 
obtained from the IR reddening maps. 380 of the 402 gE galaxies have 2MASS photometry.
Also shown are the 17 galaxies in the Norma
cluster for which we have obtained IRSF photometry so far (filled circles),
and those galaxies in our sample for which 2MASS photometry is available
($N = 51$, open circles). The mean extinction-corrected ($J^{20}-K^{20}_s$)$^0$ 
colour of the 380 gE galaxies is $0.894 \pm 0.003$ mag (illustrated by the
horizontal dotted line). The solid line corresponds to $E(J-K_s) = 0.52 E(B-V)$,
assuming a standard extinction law with $R_V = 3.1$ (Cardelli et al.~1989).

The mean $J^{20}-K^{20}_s$ colour of the 16 galaxies with accurate
IRSF photometry is $1.042 \pm 0.013$ mag (rejecting the outlying galaxy 
(WKK\,6299) due to contamination of a very bright nearby star). This means
that from our photometry we deduce a mean value of $E(B-V) = 0.22 \pm 0.03$ 
mag\footnote{Note that we have applied a correction for the change in 
the spectral energy distribution of a redshifted  (and reddened) elliptical galaxy 
(McCall \& Armour 2000; Fingerhut et al.~2003).}, 
compared to $E(B-V) = 0.207 \pm 0.003$ mag derived from the IR reddening maps.
The implied difference in $A_{K_s}$ between the IR reddening maps and 
the reddening derived from our photometry, is less than 0.01 mag. This confirms that
for $E(B-V) \le 0.25$ mag, the IR reddening maps are reasonably accurate
(see also Burstein 2003; Burstein, these proceedings), although this clearly needs to be confirmed on a 
much wider scale.

\subsection{{Substructure}}

There are some indications that substructure/subclustering exists within
the rich massive Norma cluster.
X-ray observations (B\"ohringer et al.~1996) reported evidence for
an X-ray subgroup close to the core of the cluster, while a dynamical analysis
(e.g., using the Dressler-Shectman $\delta$-test; Dressler \& Shectman 1988) 
of the cluster based on 
296 redshift-confirmed cluster members within the Abell radius ($= 1.75^{\circ}$) 
shows the presence of a spiral-rich subgroup $\sim 1.2^{\circ}-1.8^{\circ}$ away from
the cluster core (Woudt 1998; Woudt et al., in prep). The latter might have had
some influence on the large peculiar velocity derived by Mould et al.~(1991).

Contamination in the FP analysis by subgroup members could result in a systematic
offset. However, with a sample of 76 galaxies, and a thorough understanding of the
dynamics of the Norma cluster, this effect will be limited, if not 
completely eliminated.

\subsection{A first look at the {Fundamental Plane of the Norma cluster}}

After the inventory of possible sources of systematic uncertainty and the 
quantitative analysis thereof, we can now turn our attention back to the original
objective of this study: the near-infrared FP of the Norma cluster.

\begin{figure}[!ht]
\plotfiddle{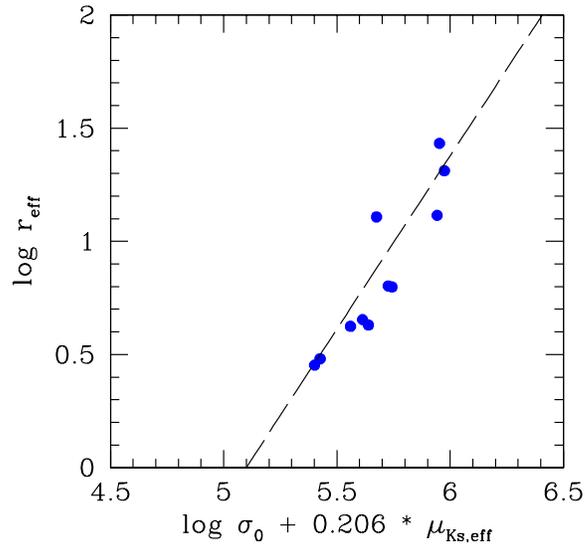}{7.0cm}{0}{65}{65}{-145}{-120}
\caption{The $K_s$-band Fundamental Plane of the Norma cluster based on 11 galaxies. The dashed
line is not a fit to the data, but corresponds to the FP of the Perseus cluster (Pahre, Djorgovski
\& de Carvalho 1998) which has a similar redshift to the Norma cluster; $v_{\rm CMB}$(Perseus) =
5169 km s$^{-1}$ versus $v_{\rm CMB}$(Norma) = 4928 km s$^{-1}$ (Woudt 1998).}
\label{woudtf4}
\end{figure}

We have determined so far the photometric properties ($r_{\rm eff}$, $\mu_{\rm eff}$) of 11 of the 76 galaxies.
Combined with the 2dF spectroscopic measurement of  $\sigma_0$ (the velocity dispersion), 
the FP of the Norma cluster is emerging (see Fig.~\ref{woudtf4}). 
Note that the photometric parameters have not yet been corrected for effects of seeing
(the mean seeing of our IRSF observations is $\sim 1.35''$) and so this diagram should not be regarded
as final. However, this preliminary analysis indicates that the Norma cluster has a well-defined
FP and that this cluster's peculiar velocity (in the Cosmic Microwave Background restframe) is small.
Note that the dashed line in Fig.~\ref{woudtf4} is not a fit to the data
but shows a comparison with the Perseus cluster (Pahre, Djorgovski \& de Carvalho 1998) which
has a similar redshift to the Norma cluster.

\section{Discussion}

Following careful considerations of issues regarding the Galactic foreground extinction, star-crowding,
and substructure within the Norma cluster, an accurate distance determination (accurate to $\sim 150$ 
km s$^{-1}$) to the Norma cluster will follow from our near-infrared FP analysis of 
76 early-type galaxies. 

It appears that for low-latitude clusters subjected to a modest amount of extinction and star-crowding, the
near-infrared FP can be successfully used to arrive at unbiased distance estimates. 
It will therefore be of great interest and importance to select suitable clusters for
near-infrared FP studies to supplement, for example, the 6dF $v$-survey 
(Jones et al., these proceedings) at low Galactic latitudes ($|b| \le 10^{\circ}$). Of particular 
interest are clusters in and beyond the {Great Attractor} region, in order to weigh the relative 
contribution of the GA and the {Shapley Concentration} (see also Lucey et al.
and Ebeling et al., these proceedings).

\acknowledgments{PAW and APF kindly acknowledge financial support from the National Research
Foundation. RCKK thanks CONACyT for their support (research grant 40094F) and the Australian
Telescope National Facility (CSIRO) for their hospitality during her sabbatical. 
MLM thanks the Natural Sciences and Engineering Counsel of Canada for its continuing support.
This research
has made use of the NASA/IPAC Infrared Science Archive (2MASS) and the NASA/IPAC Extragalactic
Database (NED), which are operated by the Jet Propulsion Laboratory, California Institute of
Technology, under contract with the National Aeronautics and Space Administration.}

\end{document}